\documentclass[11pt,twoside]{article}
\usepackage{graphicx,epsfig,natbib,epstopdf}
\usepackage{CS18}
\begin{document}
%
%
%
\title{Transit polarimetry of exoplanetary system HD189733 }
%
%
\author{Kostogryz, N.M.$^{1,3}$, Berdyugina, S.V.$^{1,2}$, Yakobchuk, T.M.$^{3}$}
\affil{$^1$Kiepenheuer Institut f\"{u}r Sonnenphysik, Sch\"{o}neckstr. 6,  79104 Freiburg, Germany}
\affil{$^2$NASA Astrobiology Institute, Institute for Astronomy, University of Hawaii, USA\\}
\affil{$^3$Main Astronomical Observatory of NAS of Ukraine, Zabolotnoho str. 27, 03680 Kyiv, Ukraine}

\begin{abstract}
%
We present and discuss a polarimetric effect caused by a planet transiting the stellar disk  thus breaking the symmetry of the light distribution and resulting in linear polarization of the partially eclipsed star. Estimates of this effect for transiting planets have been made only recently. In particular, we demonstrate that the maximum polarization during transits depends strongly on the centre-to-limb variation of the linear polarization of the host star. However, observational and theoretical studies of the limb polarization have largely concentrated on the Sun. Here we solve the radiative transfer problem for polarized light and calculate the centre-to-limb polarization for one of the brightest transiting planet host HD189733 taking into account various opacities. Using that we simulate the transit effect and estimate the variations
of the flux and the linear polarization  for HD189733 during the event. As the spots on the stellar disk also break the limb polarization symmetry
we simulate the flux and polarization variation due to the spots on the stellar disk. 
\end{abstract}
%
%
\section{Introduction}
HD189733 is currently the brightest ($m_V~=~7.67 mag$) star known to harbour
a transiting exoplanet \citep{2005A&A...444L..15B}. This, along with the short period (2.2 days)
and large ratio of planet-to-star radii ($R_{pl}/R_\star \approx 0.15$),
makes it very suitable for different types of observations including polarimetry
\citep{2008ApJ...673L..83B, 2011ApJ...728L...6B}.  
In this paper we model the polarimetric effect that occurs during the planetary transit
in front of the stellar disk. The planetary transit breaks the symmetry of the distribution of light coming from the stellar disk
which results in the net linear polarization. This effect was simulated first by \citet{2005ApJ...635..570C} for 
artificial exoplanetary systems who revealed that the accuracy of modern polarimeters of $1\times 10^{-6}$
is enough to detect it.
Later, \citet{2011MNRAS.415..695K} and \citet{2012AASP....2..146F} calculated the planetary transit effect for HD189733
with Monte Carlo simulations taking into account two different types of center-to-limb variations of linear polarization (CLVP)
of the host star as initial approximations such as the CLVP for pure scattering atmosphere \citep{1950ratr.book.....C} and the solar CLVP \citep{2009ApJ...694.1364T}. They showed that result of such simulation strongly depends on CLVP of 
the stellar disk. 

So far, the efforts to study the CLVP are concentrated only for the Sun (e.g. \citet{2005A&A...429..713S}, etc...) while
there is no information about the CLVP for cooler dwarf stars.
In this paper, we present our calculations of the CLVI and CLVP 
which are obtained by solving the radiative transfer equations for polarized light taking into account various opacities.
Using results of these calculations,   
we simulate polarization resulting from the planetary transit and 
stellar spots in HD189733. In Section 2, we describe our method of calculation of the CLVI and CLVP for HD189733 and in Section 3 we show results of our transit polarization simulations.
Section 4 presents discussion of our results.

\section{Stellar center-to-limb variation of intensity and linear polarization}

We solve the radiative transfer equations for polarized radiation iteratively 
assuming no magnetic field and considering a plane-parallel model atmosphere (see \citet{2014AA}).

\begin{figure}[h]
 \includegraphics[scale=0.5]{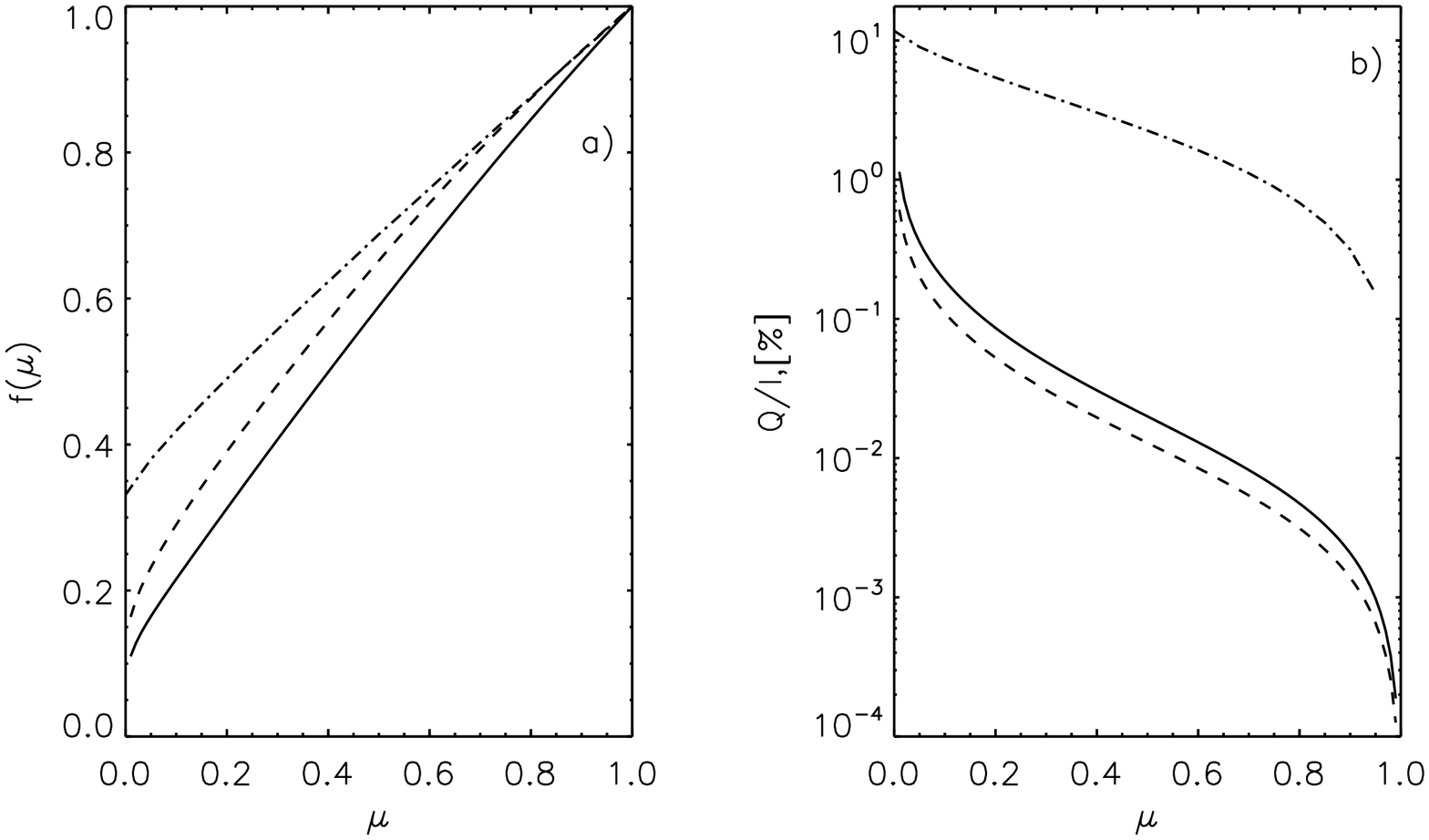}
\caption{Center-to-limb variations of (a) the intensity and (b) the continuum polarization in log scale for HD189733
(solid line), for the Sun (dashed line) and for a pure scattering atmosphere (dash dotted line) at $4500 \AA$. }
\label{Fig_clv}
\end{figure}

As the stellar model atmosphere of HD189733 we use the PHOENIX model \citep{1999ApJ...512..377H} with  
effective temperature $T_{\rm{eff}} = 5000 K$ and gravity $\log g = 4.5$. Scattering and absorption 
continuum opacities are calculated using the SLOC code \citep{1991BCrAO..83...89B} 
at the wavelength of 4500\AA.  We take into account the following contributors: 
\begin{itemize}
\item scattering opacity: 
Thomson scattering on free electrons and
Rayleigh scattering on $\rm{HI, HeI, H_2, CO, H_2O}$, and other
molecules,  
\item absorption opacity: free-free and bound-free transitions
in $\rm{H^-, HI, HeI, He^-, H_2^-, H_2^+}$, and metal photoionization.
\end{itemize}
With these contributors, the calculated CLVI and CLVP 
in the continuum spectrum of HD189733
are presented in Fig. \ref{Fig_clv}. In addition, we consider the solar CLVI  and CLVP \citep{2014AA} and 
the CLVs calculated by \citet{1950ratr.book.....C} for pure scattering atmosphere. 

As is seen in Fig.\ref{Fig_clv}, the continuum polarization for HD189733 is 
slightly larger than for the Sun, while it is significantly smaller than the polarization 
calculated in \citet{1950ratr.book.....C} for pure scattering atmosphere.
We conclude that neither Chandrasekhar's calculation nor the solar CLVP 
can reproduce correctly the CLV of the polarization for the K dwarf star HD189733.

\section{Polarization of the transiting system}

\begin{figure}[h]
 \includegraphics[scale=0.45, angle=-90]{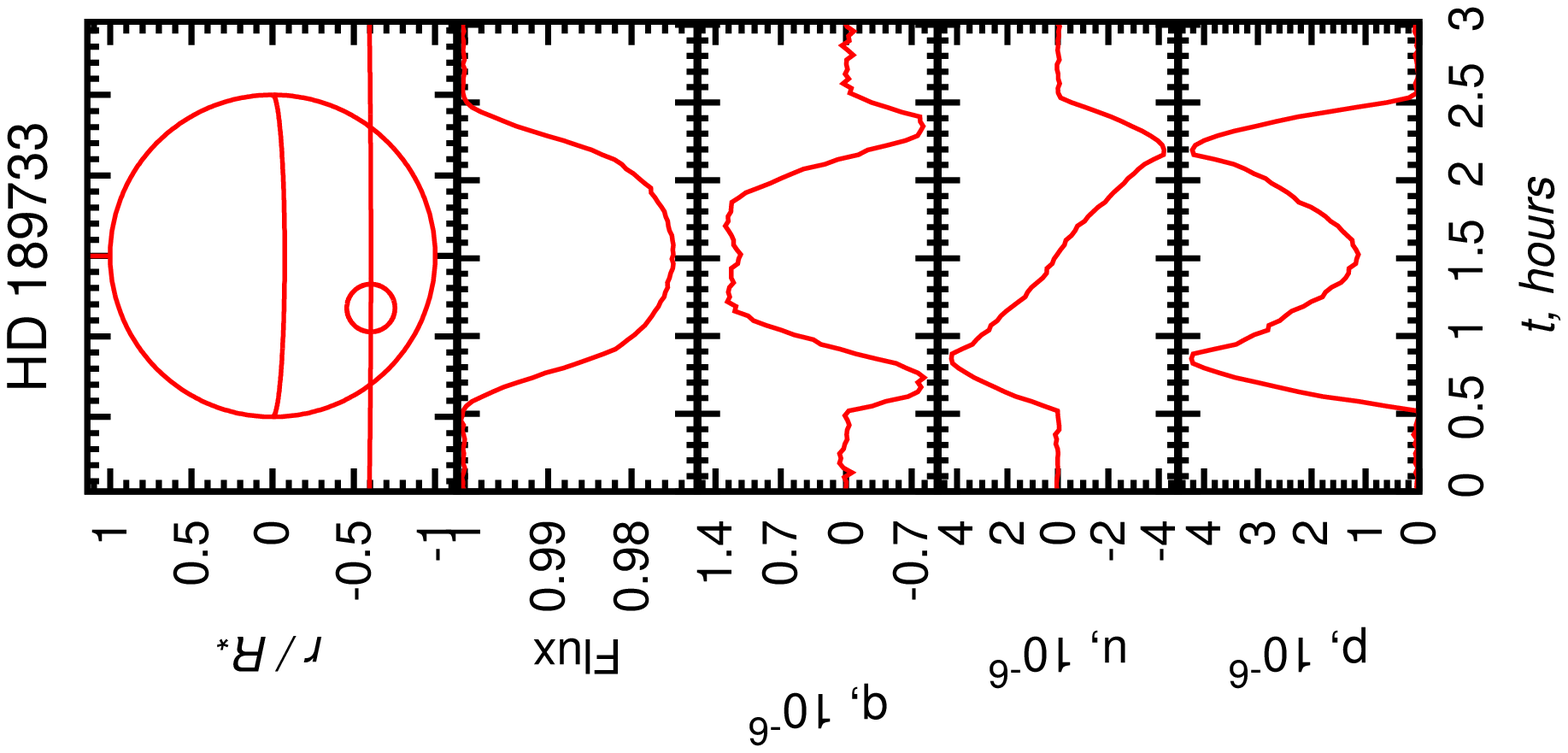}
\includegraphics[scale=0.45, angle=-90]{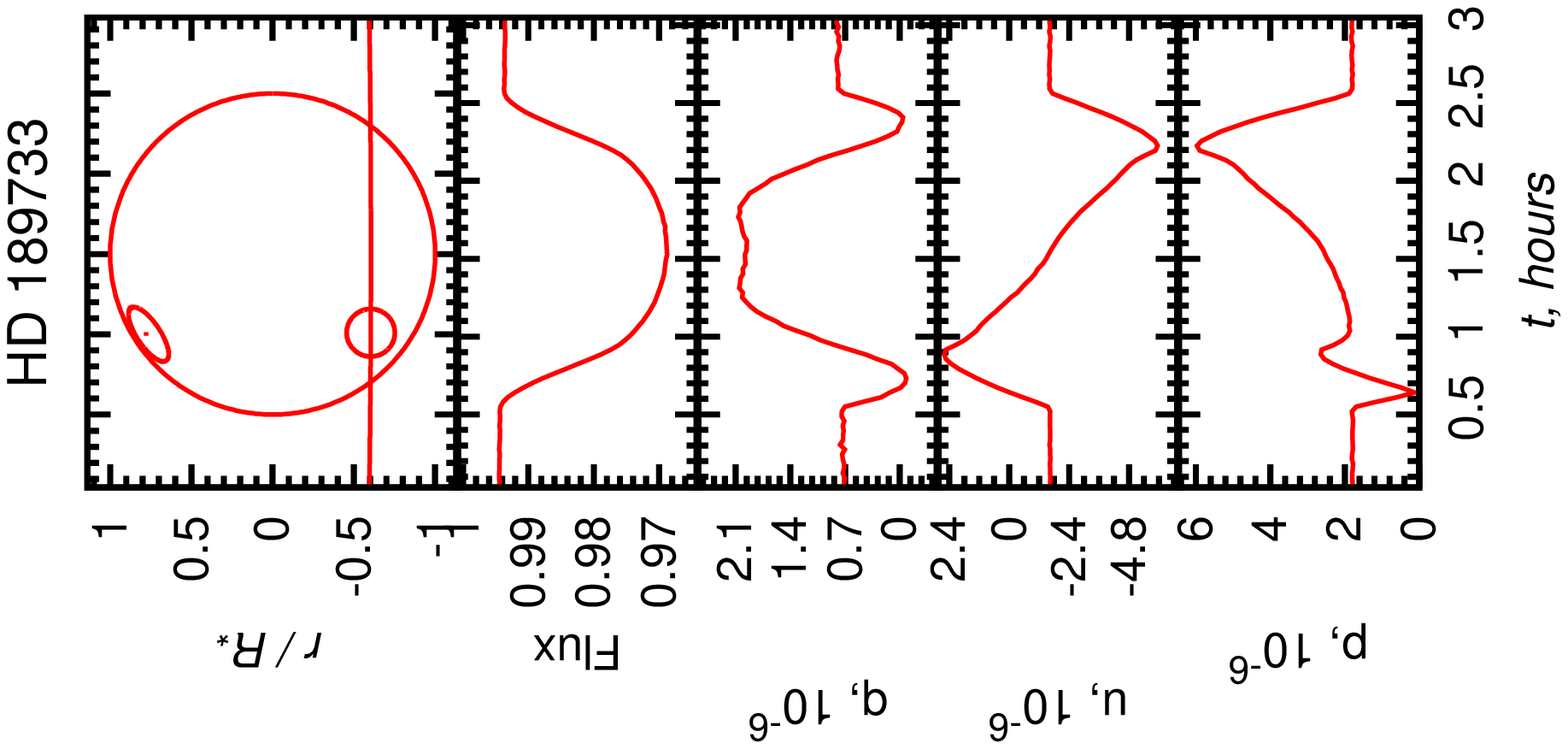}
\includegraphics[scale=0.45, angle=-90]{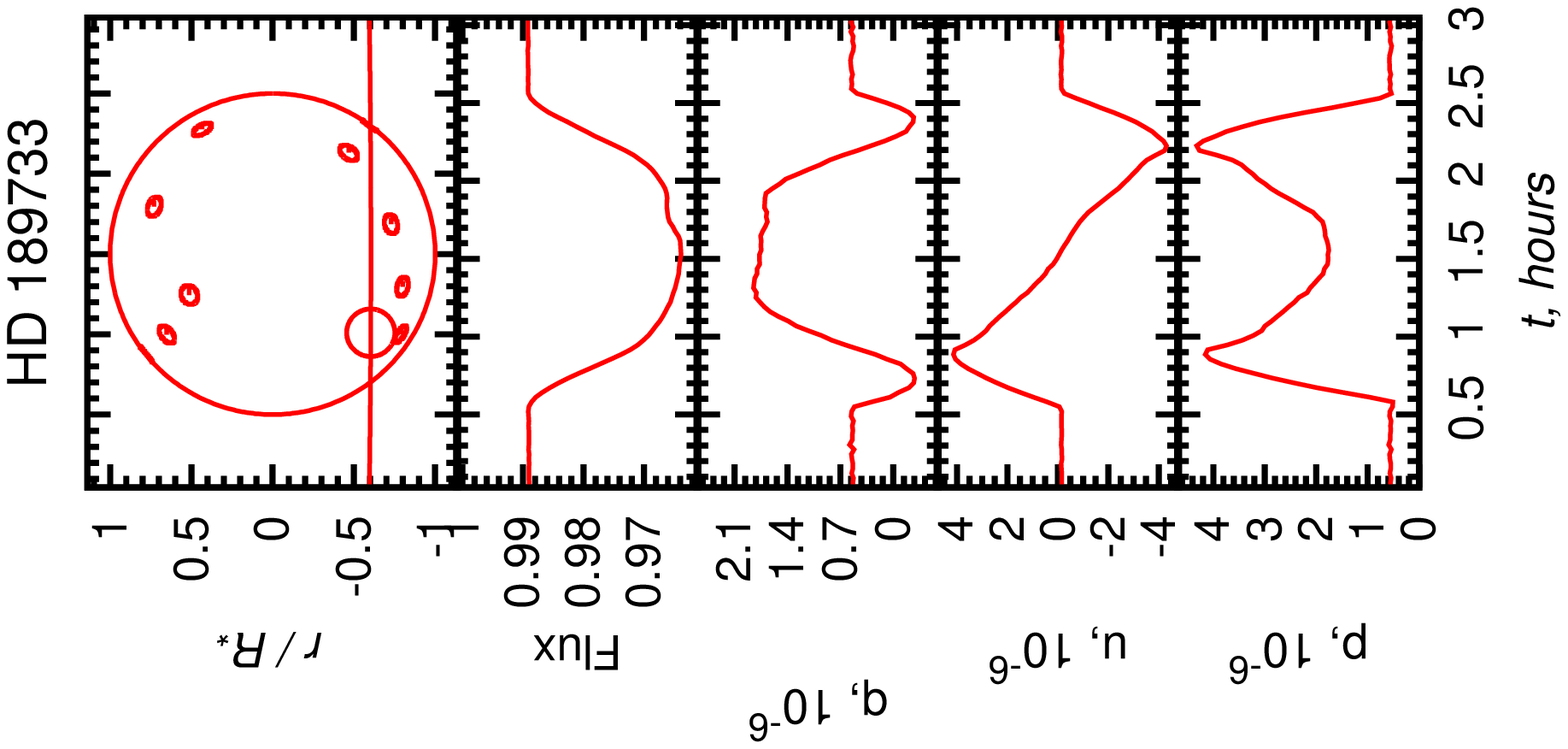}
\caption{Simulations for the HD189733 system. Each column top to bottom contains panels transit(starspot) geometry, the flux and the polarization variations (normalized 
Stokes parameters $q, u$, and polarization degree p)
during the planet transit. The first column describes only the planetary transit, while 
the next two columns show the variation of intensity and polarization during the planetary transit
at the presence of spots on the stellar disk.}
\label{Fig_poldegree}
\end{figure}

Using Monte Carlo simulations following \citet{2005ApJ...635..570C} and \citet{2011MNRAS.415..695K}
and taking into account the CLVI and CLVP calculated for HD189733 as described in Section 2, 
we obtain the variation of the flux, the normalized Stokes $q$ and $u$ parameters and the polarization
degree during the planetary transit (Fig. \ref{Fig_poldegree}, first column). The flux and polarization curves are sensitive to orbit inclination
angle and ratio of planet-to-star radii. The flux provides the planet-to-star radii ratio and the inclination angle, while polarimetry 
provides also the information in which South or North hemisphere the planet transits the star. The value and the sign of Stokes q and u    
describe the position of the planet on the stellar disk projected onto the sky plane. As is seen in Fig. \ref{Fig_poldegree} (first column) 
the variation of the polarization degree is symmetric and 
the maxima are slightly larger than $4\times10^{-6}$ at 4500 \AA. 

Another effect that breaks the symmetry of the host star disk and results in linear polarization 
is the presence of spots on the stellar disk. 
As was shown by \citet{2005ApJ...631.1215W}, 
HD189733 is an active star and spots can cover up to 1\% of the total stellar surface. 
In our simulation we set different parameters of
spots such as temperature ($\rm{3500 K - 4000 K}$), sizes (up to 1\% of the stellar surface) 
and  locations (chosen randomly in the range of latitudes of $\pm20\deg$ to $\pm60\deg$). 
The CLVI and CLVP for spots are calculated for Phoenix model atmosphere with adopted effective temperatures and gravities.
Thus, we calculate the flux and polarization variations 
contributing from stellar activity which are shown in Fig. \ref{Fig_poldegree} (second and third columns). 
In both columns for times before and after the  planet transit, we obtain a flux decrease and a polarization increase caused
only by spots present on the stellar disk. The variation of the polarization provides information about the position of starspots.
Since the polarization of the star is the largest at the limb, Fig. \ref{Fig_poldegree} (second column) shows simulation for the extreme 
case of the biggest possible spot at the limb for HD189733. In this case, the polarization due to the spot is the largest polarization degree $\approx 2\times 10^{-6}$. \citet{2011ApJ...728L...6B} assumed the same extreme starspot parameters and estimated
the maximum polarization from spot up to $3\times10^{-6}$ which is in very good agreement with our simulations.

 The third column of the Fig. \ref{Fig_poldegree} shows the same parameters as the other two columns but for another distribution of starspots. In this case we distribute 8 spots more or less homogeneously over the stellar disk between the latitudes $\pm20\deg$ to 
 $\pm60\deg$ with the total area covered by spots of about 1\% of the stellar disk. It is seen, that the cases of the spotless star and the star with the distributed spots lead to very similar signals in all panels which mostly differ by vertical offsets 
 of $0.5\times10^{-6}$ which corresponds to the signal from the spots. We would like to draw your attention that in Fig.\ref{Fig_poldegree} (third column) at t=1.7h the planet covers one of the spots which leads to little bumps in all the curves below.

\section{Summary}
In this paper, we calculate the center-to-limb variations of the intensity and linear polarization in stellar continuous for HD189733. 
Similar to limb darkening measurements in transit photometry, we show that it is possible to measure the center-to-limb variations of the linear polarization
by means of transit polarimetry. Transit polarimetry provides information about the ratio of planet-to-star radii, orbit inclination and the transit hemisphere. In addition, it constrains the sizes and positions of starspots
on the stellar disk, while transit photometry cannot, unless the planet covers a spot during its transit.

Using Monte Carlo simulations we also obtain the linear polarization for the transiting exoplanetary system HD189733 with the found maximum equal to $4.4 \times 10^{-6}$ at 4500\AA. In order to measure this effect for HD189733 we need a very sensitive polarimeter.
 
We show that the maximum polarization
effect due to starspots is still less than $2 \times 10^{-6}$ for a single spot seen at the stellar limb and even smaller for a more homogeneous distribution of spots over the surface. Thus, the influence of the spots on the total polarization for HD189733 is too small to be measurable. It requires many planetary transit observations in order to achieve better statistics and polarimetric accuracy. 

\acknowledgments{
 This work supported by the HotMol project of the ERC Advanced Grant. We thank Martin K\"{u}rster for
 the comments that improve this paper. 
}

\end{document}